# MODELING OF ION BEAM INDUCED CHARGE SHARING EXPERIMENTS FOR THE DESIGN OF HIGH RESOLUTION POSITION SENSITIVE DETECTORS

J. Forneris[1], D. N. Jamieson[2], G. Giacomini[3], C. Yang[2], E. Vittone[1]

[1]Università di Torino, Dipartimento di Fisica e Centro di Eccellenza NIS; INFN - sez. Torino; CNISM – sez. Torino, via P.Giuria 1, 10125 Torino, Italy.

[2]Australian Research Council Centre of Excellence for Quantum Computation and Communication Technology, School of Physics, University of Melbourne, Parkville VIC 3010, Australia

[3]Fondazione Bruno Kessler, FBK-CMM, Via Sommarive 18, I-38123, Trento, Italy.

## Abstract

In a multi-electrode device, the motion of free charge carriers generated by ionizing radiation induces currents on all the electrodes surrounding the active region [1]. The amount of charge induced in each sensitive electrode is a function of the device geometry, the transport parameters and the generation profile. Hence this charge sharing effect allows the signal from each sensitive electrode to provide information about the electrical characteristics of the device, as well as information on the location and the profile of each ionization track.

The effectiveness of such approach was recently demonstrated in Ion Beam Induced Charge (IBIC) experiments carried out using a 2 MeV He microbeam scanning over a sub-100 μm scale silicon device, where the ion strike location point was evaluated through a comparative analysis of the charge induced in two independent surface electrodes coupled to independent data acquisition systems [2].

In this report, we show that the Monte Carlo method [3] can be efficiently exploited to simulate this IBIC experiment and to model the experimental data, shedding light on the role played by carrier diffusion, electronic noise and ion beam spot size on the induction of charge in the sensitive electrodes. Moreover, the Monte Carlo method shows that information on the ion strike position can be obtained from the charge signals from the sensitive electrodes.

# 1. Introduction

Ion beam induced charge is a widespread technique used since 1993 to characterize electronic materials and devices. Many previous papers have underlined the potential of this technique to investigate with micron resolution both qualitatively and quantitatively the transport properties of semiconductors [4], the damage induced by focused ion beams and, simultaneously, the effect of damage on the carrier transport properties [5] and the analysis of single event effects in microelectronic devices [6].

An important feature which makes IBIC particularly attractive for material science is the availability of a robust theoretical background [4,7], which allows the extraction of practically all the information needed to qualify materials from the electronic point of view and to analyse the performance of electronic devices in a radiation environment.

Recently, IBIC has acquired further interest for the measurement of the ion strike location in multielectrode devices. Examples include the development of single atom deterministic doping techniques, which are now emerging [2,8,9,10] for potential applications in solid state quantum information processing devices and addressing issues with variations in classical device characteristics arising from statistical doping. For such purposes the evaluation of the induced charge shared between the electrodes could be used for the measurement of the ion strike location to sub-micron precision. The achievement of this goal requires accurate analysis of the induced charge pulse formation mechanism and a suitable investigation of the main physical phenomena influencing the measurement.

In this paper, we present a model based on the Shockley-Ramo-Gunn theorem [4, 7] and developed through a Monte Carlo approach [3]. We apply this model to an experiment recently reported by Jong et al. [2] which aimed to investigate the charge sharing phenomena occurring in a multi-electrode silicon p-i-n device.

The motivation of such investigation is firstly to validate the model in a charge sharing configuration through a direct comparison of the Monte Carlo simulation with the experimental results; secondly to extract information on the role played by electrons and holes in the induced charge signal formation and on the effects of electronic noise and microbeam spot size on the definition of the Charge Collection Efficiency (CCE) profiles; finally, to assess the suitability of the method based on the measurement of the charge shared between the electrodes to estimate the ion strike position at the submicron level.

The simulated IBIC experiment is described in detail in previous papers [2, 8]. For ease of reading, we summarize in Fig. 1 the main features of the device and the results of the IBIC experiment.

Fig. 1a shows the device geometry, the doping distribution and the electrical connections: the two sensitive electrodes located at the irradiated surface were reverse biased at -20 V with respect to the grounded back electrode and connected to two independent charge sensitive electronic chains.

The IBIC experiment was carried out using a 2 MeV $He^+$ microbeam for raster scanning the region between the left and right top electrodes (to which we will refer as L and R in the following). The ionization map relevant to this microbeam, which will be used in the simulation, was evaluated by the SRIM2010 code [11] and is shown in Fig. 1b.

The median CCE profiles acquired by L and R electrodes (Fig. 1c) arise from the sharing of charge induced by ion strikes in the ≈5 μm wide region centred at the point of symmetry of the system $x_0$ (i.e., symmetry axis of the device). The CCE pulse pairs recorded from L and R sum to the nominal energy of the ion beam and allow for the spatial calibration of IBIC signal, leading to the identification of the ion strike location from the signal amplitude on both electrodes [2].

## 2. Modelling

The numerical simulation of the IBIC experiment is based on the Shockley-Ramo-Gunn theory [4, 7]. This model incorporates the role of charge carriers and the dependence of CCE profiles on physical properties (geometry of the system, electrostatic configuration, transport parameters) and the experimental device attributes (electronic noise, threshold level).

The model is based on the numerical, finite element evaluation of the solution of Poisson's equation for the electrostatics of the device. The charge collection efficiency was then mapped using a recently developed Monte-Carlo approach to the solution of the drift-diffusion equation for charge carriers [2], following the prescriptions of Gunn's theorem for the induced charge.

### 2.1 Electrostatics

An electrostatic model of the device was simulated, according to geometry, doping concentrations and bias voltages (-20 V at the top electrodes) defined in Fig. 1a, by means of commercial Technology Computer Aided Design (TCAD) packages [12, 13]. The electric potential from this model is depicted in Fig. 2a, together with electric field streamlines. The device is fully depleted, with the electric field being non-zero over the whole region between top and back electrodes. Due to the presence of two electrodes on the top surface, the electric field splits around $x_0$, where the vertical component vanishes and the field lines are defined by a non-negligible contribution in the horizontal direction.

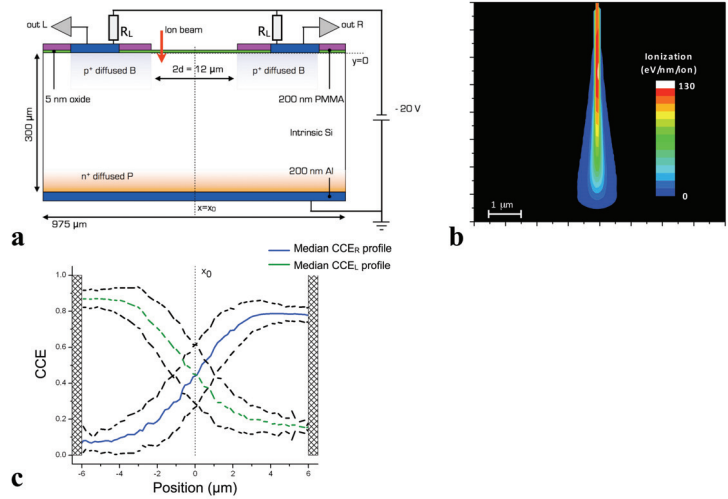

**Fig. 1:** (a) Detector geometry in cross section; (b) 2D ionization profile of 2 MeV He ions in Si from SRIM simulation. (c) Experimental median CCE profiles collected from L and R electrodes. Dashed lines represent the dispersion (1st and 3rd quartile) of the CCE distribution (from Jong et al. [2]).

The calculated weighting potential associated with L is presented in Fig. 2b (a mirror image gives the results for R). It is worth noting that the weighting potential is non-zero in the right semi plane ($x>x_0$), meaning that a charged particle moving in a neighbourhood of R induces a signal on L as well.

Fig. 2c shows the vertical component of the electric field and the weighting potential profiles along the vertical line at $x=x_0$. The weighting potential profile relevant to L exhibits a monotonically decrease from the value at the top surface (0.5, being $x_0$ at half distance between electrodes). On the other hand, the vertical component of the electric field has a similar monotonic behaviour in the whole device except in the first few microns in depth, where its contribution is almost zero. The proportion of charge carriers that drift towards L or R is due to the horizontal component of the electric field which is presented, at y=0, in Fig. 2d together with the line plot of the weighting potential for L.

The horizontal component of the electric field displays a node at $x_0$, where the sink for the electric field lines changes from L to R at increasing x. It is worth noticing that in the first few microns around $x_0$ the carrier velocities are far from saturation and display a linear profile. Along the horizontal axis the L weighting potential exhibits a decreasing behaviour from 1 (L) to 0 (R), with median value at $x_0$ (0.5).

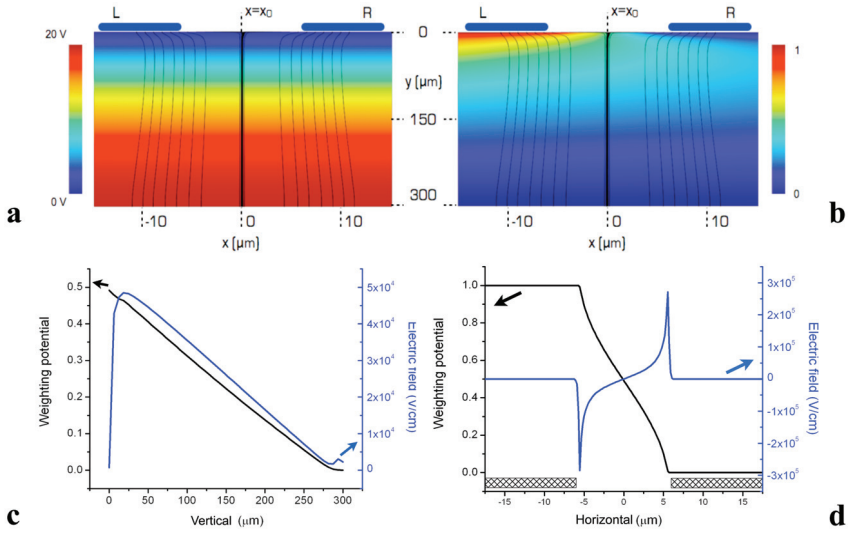

**Fig. 2:** Electrostatic potential (a) and weighting potential map associated with L electrode (b) with electric field streamlines. (c) weighting potential and vertical component of the electric field along the vertical line at x=$x_0$; (d) left electrode weighting potential and horizontal component of the electric field along the horizontal line at y=0; the horizontal bars indicate the electrodes.

## 2.2 Monte Carlo simulation

The electric field within the device was used as input for a 2D Monte Carlo simulation of the IBIC experiment [3]. In order to be able to use the IBIC signals from L and R for identifying the ion strike location with sub-micron precision, a fine square mesh grid with spatial step size Δx=50 nm was considered for the model calculations, leading to a time step of $\Delta t_n$=1.5·10$^{-13}$ s, $\Delta t_p$=5·10$^{-13}$ s for electrons and holes respectively [14]. The field dependent expression of carrier mobility from Caughey and Thomas [15] was used and the relevant diffusivity values were defined through the Einstein relationship [16].

Simulation was performed as follows:

(1) Each ion strike was assumed to induce 1000 e-h pairs in the device, randomly generated according to the SRIM 2D ionization distribution, which takes into account for lateral straggling of the ion along its path in silicon, centred at the nominal position of incidence of the microbeam (see Fig. 1b).

(2) For each generated carrier, a drift-diffusion random walk was simulated according to prescriptions in [3], setting a mean carrier lifetime of 1 µs. This value is supposed not to significantly influence the pulse formation mechanism since in the high electric field region the carriers drift time is of the order of ns; the extension of the low electrical field region (in the neighbourhood of $x_0$) is of the order 2 μm which is significantly smaller than the carriers diffusion length.

(3) The charge $q_j$ induced by the j-th carrier was then evaluated according to the expression

$$(1) \quad q_j = q \cdot \left( \frac{\partial \psi(x_f, y_f)}{\partial V} - \frac{\partial \psi(x_i, y_i)}{\partial V} \right)$$

where q is the elementary charge, ψ is the electrostatic potential, V is the applied bias voltage at the sensitive electrode, $\frac{\partial \psi(x, y)}{\partial V}$ is the Gunn's weighting potential and the subscripts i, f refer to the initial and final position, respectively.

(4) The Monte Carlo simulation was then carried out by following the random walk of a predefined population of carriers generated at different points along the ion tracks, evaluating their position after N time steps and hence calculating the induced charge through eq. (1).

The CCE was obtained by averaging over the induced charge of each carrier; the total CCE for each ion was finally given by the sum of electron and hole average contributions. The microbeam raster scan was finally simulated by the evaluation of the CCE of 100 ions for each nominal generation point between L and R along the horizontal direction with a constant step of 50 nm.

## 2.3 Shared charge pulse formation

In order to evaluate the contribution of different phenomena occurring in charge pulse formation, we consider an ideal experiment with a delta-like ion beam and an ideal amplification chain, with no electronic noise. Fig. 3a shows the simulated CCE profiles acquired from the two (L and R) sensitive electrodes which results from the sum of the electron (Fig. 3b) and hole (Fig. 3c) contributions.

Since e-h pair generation occurs along the ion track down to a few microns below the top surface, electrons drift along a vertical path towards the anode (back electrode), where the value of weighting potentials for both the sensitive electrodes is zero (see Fig. 2b). The CCE value, as a function of the striking point, will therefore correspond to the convolution of the weighting potentials (both for R and L) with the generation profile.

The interpretation of the hole contribution to the induced charge pulse formation, considering for example electrode L, can be interpreted as follows. Since e-h pairs generation occurs within 7-8 μm in

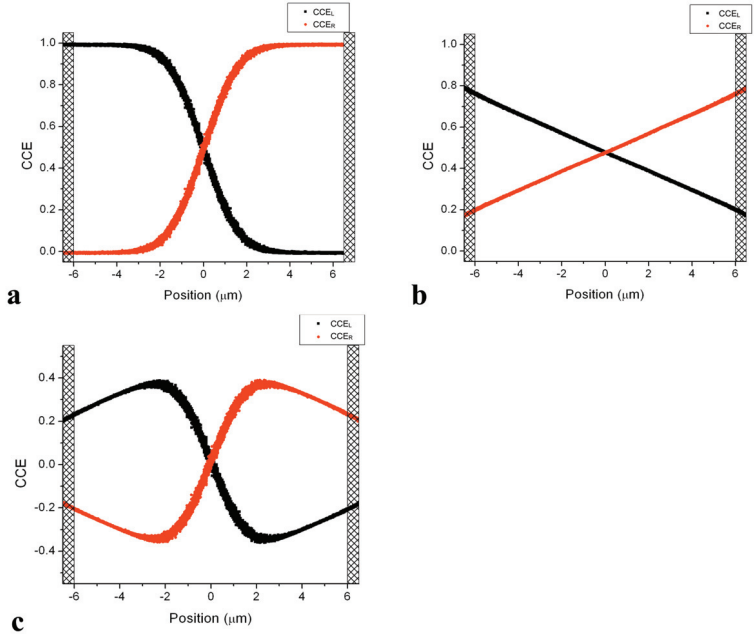

**Fig. 3** (a) simulated CCE profiles collected from left (L) and right (R) sensitive electrodes; (b) electron contribution; (c) hole contribution. Each marker corresponds to the contribution of a single ion hit (1000 simulated e-h pairs). The simulation is relevant to a delta-like ion beam; the electronic noise and threshold were not considered.

depth, hole drift in the region between L and R is practically unaffected by the vertical component of the electric field (see Fig. 2c). Therefore, motion is primarily ruled by the horizontal component of the field (Fig. 2d), which exhibits a node at $x_0$.

Holes generated at $x < x_0$ then drift towards L, where the weighting potential for L is equal to 1. Therefore, according to eq. (1), the left CCE increases with the distance from L, according to

$$(2) \quad CCE_L(x < x_0) = \left(1 - \frac{\partial \psi(x)}{\partial V_L}\right)$$

On the other hand, holes generated for $x > x_0$, drift towards R, where the L weighting potential is zero. As a consequence, the weighting potential of the initial position is always larger than the weighting potential of the final position (i.e. electric field and weighting field are antiparallel [17]) and, according to eq. (1), holes provide a negative contribution, monotonically increasing to zero with the distance from $x_0$,

$$(3) \quad CCE_L(x > x_0) = -\frac{\partial \psi(x)}{\partial V_L}$$

The resulting CCE curve is anti-symmetric by reflection with respect to $x_0$.

Finally, electrons and holes contributions sum to the total CCE profile. Considering the L sensitive electrode, in the right semi plane, a negative contribution from holes and a positive contribution from electrons cancel out; on the other hand, both contributions are positive at the left side and they sum up to CCE=1. The resulting left (right) CCE curve has a step-like behaviour, softened by a sigmoidal decrease (increase) centred on $x_0$.

### 2.4 The role of the microbeam spot size

The Monte Carlo method allowed for the simulation of an experiment involving a finite microbeam spot size, corresponding to the dispersion of the actual generation position around its nominal value. Ion beam dispersion was simulated by choosing a random generation point normally distributed around the nominal microbeam position.

Fig. 4a shows the $CCE_L$ and $CCE_R$ profiles assuming a microbeam size (FWHM) equal to 2 μm. It is apparent that the ion strike point dispersion induces a broader distribution of the charge pulses when compared with Fig. 3a, which is relevant to an ideal delta-like ion beam. The broadening is clearly more apparent within 2 μm from $x_0$, whereas at larger distances the CCE saturation at 100% (see Fig. 3a) is insensitive to the microbeam size.

### 2.5 The role of the electronic noise and amplification threshold.

The effect of the electronics chain on the pulse acquisition was simulated by adding to the simulation output a normally distributed random number with zero average value and considering only CCE values higher than a nominal electronic threshold.

Fig. 4b shows the $CCE_L$ and $CCE_R$ median profiles calculated assuming a noise FWHM of 10%. Similar to the effect of variations in the ion strike point, the electronic noise broadens the pulse distributions. Differently from the effect of the beam dispersion, the broadening extends over the entire profile.

Fig. 4c shows the effect of an electronic threshold set at a value of CCE equal to 10% (corresponding to 200 keV in terms of ion energy). This threshold was set on the same data displayed in Fig. 4b. Consistent with the experimental procedure [2], ($CCE_L$,$CCE_R$) pairs were selected so that their sum

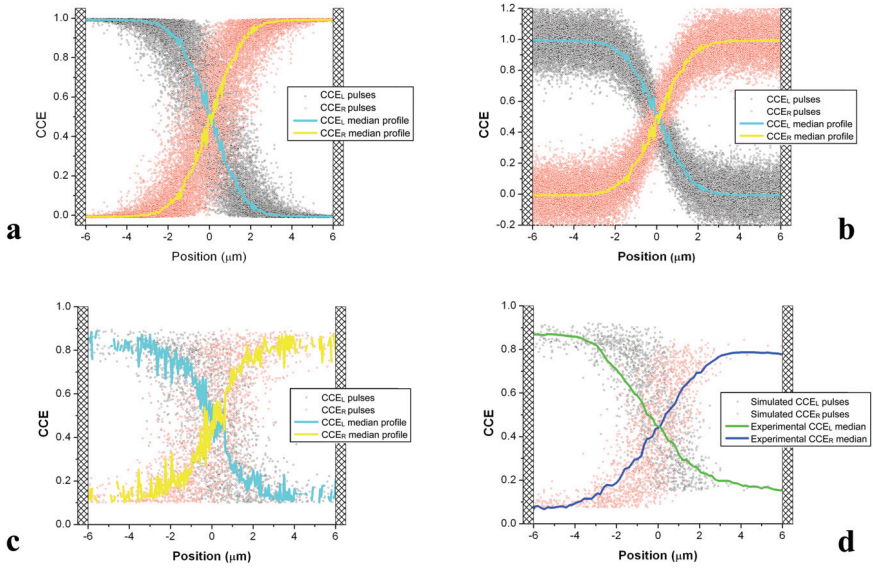

**Fig. 4** simulated CCE profiles with associated median curves collected from left (L) and right (R) sensitive electrodes assuming

(a) an ion beam with a Gaussian profile with FWHM of 2 μm without any noise from the electronic chain.

(b) a delta-like ion beam with an electronic noise with an FWHM of 10% of CCE

(c) a delta-like ion beam with an electronic noise with an FWHM of 10% and an electronic threshold of 10% of CCE

(d) simulation of the IBIC experiment described in [1] assuming a beam spot size of 2 μm (FWHM) a Gaussian noise with FWHM=10% of CCE and electronic thresholds of 15% and 7.5% for the left and right electrodes, respectively. Solid lines: experimental median profiles.

was equal to the nominal energy of the ion beam, meaning that only L-R pulse pairs so that $0.95<CCE_L+CCE_R<1.0$ were considered. With this constraint, the elimination of sub-threshold pulses acquired by one electrode implies also the exclusion of the corresponding pulses acquired by the other electrode relevant to the same ion strike position. This is the reason for the thinning-out of the signal density, particularly evident at the extreme ends of the CCE profiles.

### 2.6 Simulation of the IBIC experiment.

Fig. 4d shows the result of the simulation of the experiment reported in [2]. A single marker indicates the $CCE_L$ or $CCE_R$ values at the relevant nominal ion strike position.

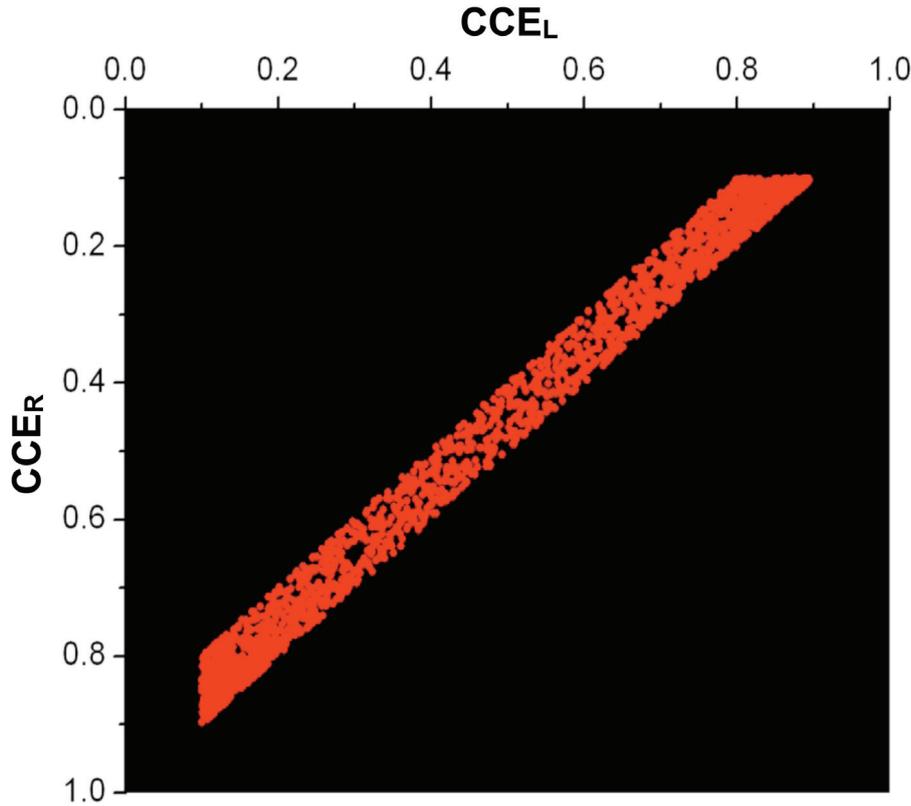

**Fig. 5:** the distribution of ($CCE_L$, $CCE_R$) ordered pairs from each simulated ion impact. To be compared with Fig. 4 in [2].

As before, the simulation was carried out assuming a microbeam spot size with FWHM of 2 μm, a Gaussian electronic noise with FWHM of 10% in terms of CCE and an electronic threshold of 15% and 7.5% in terms of CCE for the left and right electrode, respectively. For comparison purposes, the experimental median profiles from Jong et al. [2] are presented as solid lines. The simulated CCE profiles show good agreement with experimental data. The main causes of the discrepancies could be ascribed to a slightly different level of electronic noise and corresponding thresholds in the two independent chains of electronics connected to L and R.

Fig. 5 shows the distribution of the ($CCE_L$, $CCE_R$) ordered pairs from each single ion impact. The comparison with the equivalent experimental data shown in figure 4 of [2] is evident. This plot highlights the fact that the signal induced at each electrode is the complement to unity of the signal induced at the other electrode, i.e. $CCE_R = 1 - CCE_L$. As a consequence the combination of the two signals can be useful to reduce the uncertainty of the position by a factor of $2^{-1/2}$, but to determine the strike point position it is sufficient to consider the CCE evaluated from only one of the two electrodes. In order to clarify this crucial point, we have inverted the axis of Fig. 4d for the data relevant to the left electrode, and plotted the ion strike position as function of the simulated $CCE_L$, assuming a delta-like ion beam spot size. The resulting profile is shown in Fig. 6. At the extreme ends, data assume a vertical trend, i.e. the $CCE_L$ value is practically independent of the ion strike position, as already observed. However, in a 2 μm wide region centred at $x_0$, the data points gather around the regression line with slope of (-3.91±0.11) μm per unit of CCE. This means that a spectral resolution of 10% in terms of CCE, convoluted with the uncertainty of the slope evaluation, leads to a spatial resolution of about 0.6 μm, in excellent agreement with the estimation carried out by Jong et al. [2]. This uncertainty in position can be further reduced to 0.4 μm if both the signals ($CCE_L$ and $CCE_R$) are taken into account.

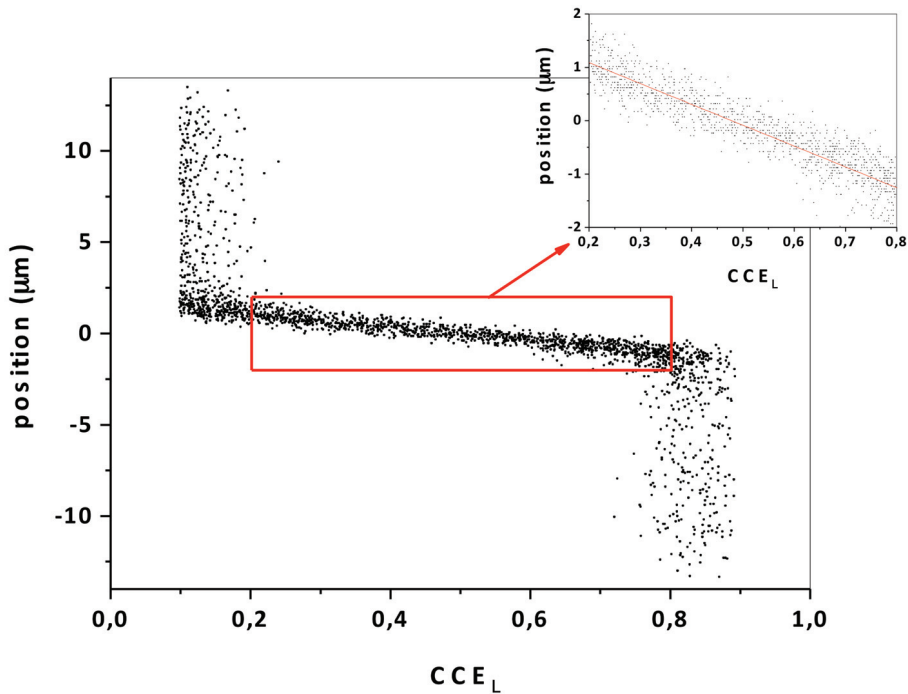

**Fig. 6:** simulated ($CCE_L$,position) events assuming a delta-like ion beam spot size. The inset shows the zoomed central region. The continuous line is the linear fit with equation
x=( 1.87±0.06)+(-3.91±0.11)·$CCE_L$.

# 3. Conclusions

This paper deals with the analysis of the induced charge signals produced by 2 MeV He ions scanning the linear gap between two surface electrodes in a p-i-n silicon junction diode. The charge induction phenomena were studied using a model based on the Shockley-Ramo-Gunn theorem and implemented by a recently developed Monte Carlo method. The simulations led to results in good agreement with the experimental data and allowed the role played by electrons and holes in the pulse formation mechanism to be modelled. In particular, it was shown that electrons provide a positive CCE contribution linearly decreasing as function of the distance from the sensitive electrode, whereas the hole contribution shows a non monotonic profile with negative values if the ions strike in the semi plane opposite to the sensitive electrode. The Monte Carlo approach was proven to be efficient to simulate the effect of electronic noise, amplification threshold, the microbeam spot size and to estimate their relevant role both in the spatial and in spectral resolution.

This work reinforces the conclusions drawn in the original report [2] about the potential use of independent IBIC signals from surface electrodes to measure the ion strike location. Moreover, a further outcome of the simulation is the evidence that the spatial discrimination with submicron resolution is possible only in a restricted region overlapping the symmetry axis. Appropriate masks could restrict ion strikes to this location where also post-implant nanocircuitry could be fabricated to exploit the implanted ions.

If we restrict the analysis to the charge sharing region at few microns beneath the surface, the vertical component of the electric field reduces and the horizontal component shows a linear trend with a node at the symmetry point ($x=x_0$); as a conclusion, transport is no more dominated by drift, but diffusion plays a non-negligible role making possible the transport in the right semi plane of carriers generated at $x<0$ and vice versa. At $y=0$, i.e. at the detector surface, the analytical expression (not shown here) of the horizontal CCE profile calculated by solving the adjoint continuity equation of holes [18], predicts that the slope of the CCE profile in the charge sharing region is a function of the ratio of the drift velocity to the diffusion coefficient. By virtue of the Einstein diffusion/mobility formula, the drift velocity increases as the temperature decreases which therefore would also increase the gradient of the of the CCE profile as a function of position in the charge sharing region. Hence operation of the device at low temperature will improve the resolution of the ion strike location measurement for the development of single atom deterministic doping techniques which use keV ions for sub-micron donor implant in silicon.

## Acknowledgements

Authors would like to thank the personnel of University of Torino - Physics Department's Computing Centre, (S. Bagnasco, D. Berzano, R. Brunetti, S. Lusso), for granting access to the newborn cloud infrastructure, on top of which Monte Carlo simulations were performed. The research by D.N. Jamieson and C. Yang was supported by the Australian Research Council Centre of Excellence for Quantum Computation and Communication Technology (project number CE110001027) and the U.S. Army Research Office (W911NF-08-1-0527). The support of R. Szymanski for the experimenal program is gratefully acknowledged.

## References

[1] W. Shockley, J. Appl. Phys. 9 (1938) 635

[2] L.M. Jong, J.N. Newnham, C. Yang, J.A. Van Donkelaar, F.E. Hudson, A.S. Dzurak, D.N. Jamieson, Nucl. Instr. and Meth. B **269** (2011) 2336

[3] P. Olivero, J. Forneris, P. Gamarra, M. Jakšić, A. Lo Giudice, C.Manfredotti, Ž. Pastuović, N. Skukan, E.Vittone, Nucl. Instr. and Meth. B **269** (2011) 2350

[4] M.B.H. Breese, E. Vittone, G. Vizkelethy, P.J. Sellin, Nucl. Instr. and Meth. B **264** (2007) 345.

[5] Ž. Pastuović, E. Vittone, I. Capan, M. Jakšić, Appl. Phys. Lett. **98** (2011) 092101

[6] G. Vizkelethy, B.L. Doyle, D.K. Brice, P.E. Dodd, M.R. Shaneyfelt, J.R. Schwank, Nucl. Instr. and Meth. B **231** (2005) 467.

[7] E.Vittone, Nucl. Instr. and Meth. B **219-220** (2004) 1043.

[8] D.N. Jamieson, C. Yang, T. Hopf, S.M. Hearne, C.I. Pakes, S. Prawer, M. Mitic, E. Gauja, S.E. Andresen, F.E. Hudson, A.S. Dzurak, R.G. Clark, Appl. Phys. Lett. **86** (2005) 202101.

[9] A. Batra, C.D. Weis, J. Reijonen, A. Persaud, T. Schenkel, S. Cabrini, C.C. Lo, J. Bokor, Appl. Phys. Lett. **91** (2007) 193502

[10] J. A. Van Donkelaar, A. D. Greentree, A. D. C. Alves, L. M. Jong, L. C. L. Hollenberg, D. N. Jamieson, New J. Phys. **12** (2010) 065016

[11] J.F. Ziegler, M.D. Ziegler, J. P. Biersack, Nucl. Instr. and Meth. B **268** (2010) 1818.

[12] ISE TCAD, available from http://www.synopsys.com .

[13] SILVACO TCAD, available from http://www.silvaco.com .

[14] B. Kaplan, Am. J. Phys. **51** (1983) 459.

[15] D.M. Caughey, R.E. Thomas, Proc. IEEE, **55** (1967) 2192.

[16] S. Selberherr, *Analysis and Simulation of Semiconductor Devices*, Springer, Vienna, 1984.

[17] V. Radeka, Ann. Rev. Nud. ParI. Sci. **38** (1988) 217,

[18] T.H. Prettyman, Nucl. Instr. and Meth. A **428** (1999) 72.